\documentclass{article}
\usepackage{spconf,amsmath,graphicx,url}

\usepackage{amssymb}
\usepackage{multirow}
\usepackage{arydshln}

\title{A Configurable Multilingual Model is All You Need to Recognize All Languages}
\name{Long Zhou$^{\dag}$, Jinyu Li$^{\ddag}$, Eric Sun$^{\ddag}$, Shujie Liu$^{\dag}$}
\address{$^{\dag}$Microsoft Research Asia \\
        $^{\ddag}$Microsoft Speech and Language Group}
\begin{document}
%
\maketitle
\begin{abstract}
Multilingual automatic speech recognition (ASR) models have shown great promise in recent years because of the simplified model training and deployment process. Conventional methods either train a universal multilingual model without taking any language information or with a 1-hot language ID (LID) vector to guide the recognition of the target language. In practice, the user can be prompted to pre-select several languages he/she can speak. The multilingual model without LID cannot well utilize the language information set by the user while the multilingual model with LID can only handle one pre-selected language. 
In this paper, we propose a novel configurable multilingual model (CMM) which is trained only once but can be configured as different models based on users' choices by extracting language-specific modules together with a universal model from the trained CMM.
Particularly, a single CMM can be deployed to any user scenario where the users can pre-select any combination of languages.   
Trained with 75K hours of transcribed anonymized Microsoft multilingual data and evaluated with 10-language test sets, the proposed CMM improves from the universal multilingual model by 26.0\%, 16.9\%, and 10.4\% relative word error reduction when the user selects 1, 2, or 3 languages, respectively.  CMM also performs significantly better on code-switching test sets.

\end{abstract}

\section{Introduction}
\label{sec:intro}

According to \cite{multilingual}, there are 40\%, 43\%, 13\%, 3\%, and less than 1\textperthousand\  people in the world can speak 1, 2, 3, 4, and more than 5 languages fluently. With the advance of deep learning \cite{DNN4ASR-hinton2012}, the commercial monolingual automatic speech recognition (ASR) systems are highly optimized with excellent recognition accuracy \cite{li2017acoustic, li2018developing}. There are increasing interests in developing high-quality commercial ASR systems that can recognize speeches from multiple languages without letting users explicitly indicate which language he/she will speak for every utterance. A common practice in industry is described in \cite{gonzalez2014real} which has an interface to enable the user to select multiple languages and use a language ID (LID) detector to select the decoding output from the ASR models of all selected languages. However, this method is cost-consuming because it needs to run multiple speech recognizers at the same time, and the LID estimation usually introduces latency because it needs a period of speech in order to have reliable decisions.

In the context of end-to-end (E2E) modeling \cite{Graves-CTCFirst, graves2012sequence, chan2016listen, chiu2018state, Li2020comparison}, the easiest way is pooling the data of all languages to build a single multilingual model. This model is a universal model, and can recognize the speech from any language as long as this language is used during training. It can be improved by taking a 1-hot LID vector as the additional input so that the multilingual model is guided to recognize that language well. The multilingual model without LID input cannot take advantage of the user selection. In contrast, the multilingual model with 1-hot LID vector needs to know which language the user will speak in advance, and cannot work for multilingual speakers who only pre-select several languages once.  Another solution is to build a specific model for every combination of languages so that we can deploy the model based on any user's selection. However, the development cost is formidable. For example, if we want to have bilingual and trilingual support of 10 languages, we have to build $C_{10}^2=45$ and $C_{10}^3=120$ specific models with such solution. 

In this work, we design \textit{a configurable multilingual model (CMM)} that can be configured to recognize speeches from any combination of languages based on user selection. We formulate the hidden output as the weighted combination of the output from a universal multilingual model and the outputs from all language-specific modules. The universal model is language independent, modeling the shared information of all languages. The residue of any language from the shared one only carries much less information. Therefore, it only needs a very small number of parameters to model the residue for every language. At runtime, the universal model together with corresponding language-specific modules are activated based on the user selection.

CMM is different from the multilingual ASR model with 1-hot LID vector which can only recognize the pre-selected single language. 
CMM also differs from the recent multilingual ASR models using mixture of experts (MoE) \cite{Das2021Multi, gaur2021mixture}, in which every expert has the same amount of parameters as the universal model. Therefore,  it is very hard for MoE to scale up with multiple languages given the very large model size. In contrast, CMM is only slightly larger than the universal model due to the residue modeling. More importantly, to our best knowledge, there is no work of configuring a single model for better recognition of any combination of languages selected by  multilingual users. 

\section{Model}
Our goal is to design a single model which can be configured at the inference time to recognize any language combination based on user selection.
This is realized with our proposed configurable multilingual model which is based on the multilingual streaming Transformer Transducer model.

\subsection{RNN and Transformer Transducer}
Because of its streaming nature, RNN-Transducer (RNN-T) \cite{graves2012sequence} has become a very promising E2E model in industry to replace the traditional hybrid models \cite{sainath2020streaming, Li2020Developing, zhang2021benchmarking}.  RNN-T contains an encoder network, a prediction network, and a joint network. 
The encoder network converts the acoustic feature $x_t$ into a high-level representation $h_t^{enc}$,  where $t$ is time index.
The prediction network produces a high-level representation $h_u^{pre}$ by conditioning on the previous non-blank target $y_{u-1}$ predicted by the RNN-T model, where $u$ is the output label index. 
The joint network is a feed-forward network that combines the encoder network output $h_t^{enc}$  and the prediction network output $h_u^{pre}$ to generate $h_{t,u}$ which is used to calculate softmax output.

Given the great success of Transformer \cite{vaswani2017attention}, 
Transformer Transducer (T-T) \cite{yeh2019transformer, zhang2020transformer} was proposed to replace LSTM with Transformer \cite{vaswani2017attention} in the encoder of Transducer with significant gain. To deal with the large latency and heavy computation cost of T-T, Chen et al. \cite{chen2020developing} proposed an efficient implementation of T-T with very small latency and computation cost, while maintaining high recognition accuracy. We use the  T-T model in \cite{chen2020developing}  as the backbone model in our study.

\subsection{Multilingual Speech Recognition}
\label{multilingual-speech-recognition}

Training a single ASR model to support multiple languages is promising and challenging \cite{toshniwal2018multilingual,kannan2019large,pratap2020massively,li2021scaling}.
Through shared learning of model parameters across languages \cite{huang2013cross, heigold2013multilingual, ghoshal2013multilingual}, multilingual ASR models can perform better than monolingual models, particularly for those languages with less data.
Besides, they significantly simplify the process of model deployment and resource management by supporting $n$ languages with a single ASR model rather than $n$ individual models.
This paper focuses on the streaming end-to-end multilingual ASR system, which predicts a distribution over the next output symbol $P(y_u|x_t, y_{u-1})$.


Previous work has demonstrated the importance of language ID (LID) \cite{li2018multi,waters2019leveraging}, with which the multilingual system can significantly outperform the universal multilingual system without LID.
A simple but effective way to leverage the LID is representing the LID as a 1-hot vector, and appending it to the input layer of the encoder network. Formally, the new input acoustic feature vector $x_{t}^{new}$ can be denoted as:
\begin{equation}
    x_{t}^{new} = [x_t; d_l]
\label{equ:concat}
\end{equation}
where $[;]$ means concatenation operation, and $d_l$ is a 1-hot vector where the corresponding dimensionality of LID is equal to one, others are zeros, for example $[0,0,0,1,0]$.

\begin{figure}[t]
  \centering
  \includegraphics[width=0.85\linewidth]{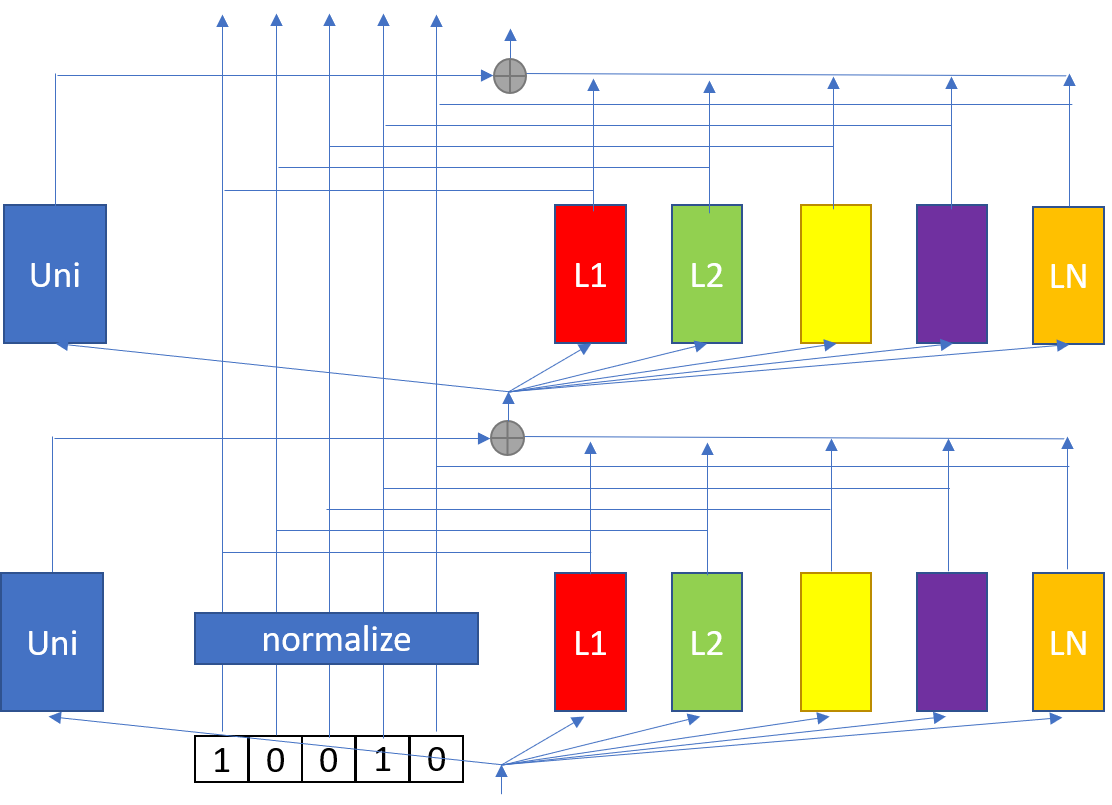}
  \caption{Diagram of configurable multilingual model (CMM). Uni denotes universal multilingual model, and Li denotes specific layer for language $L_i$.}
  \label{fig:CMAT}
\end{figure}

Although the multilingual model with the 1-hot LID vector can obtain a significant improvement than a universal model without LID by taking advantage of the user selection, it needs to know which language the user will speak in advance for every utterance, and cannot work for the popular scenario where the multilingual user can speak few languages and pre-select those languages once in the interface.

\subsection{Configurable Multilingual Model}
\label{sec:cmm}

To cover all the usage scenarios for multilingual users, we propose a configurable multilingual model (CMM) to support the scene where the utterance is from one of several user-selected languages.
Figure \ref{fig:CMAT} shows the encoder network part of CMM. The universal module (uni) is the same as the Transformer encoder of a standard multilingual ASR system. 
Compared to the universal model, CMM employs \textbf{language-specific embedding}, \textbf{language-specific layer}, and \textbf{language-specific vocabulary} to achieve the highly configurable goal.


We use the multi-hot vector as the user choice vector to represent languages selected by the user and concatenate it with input acoustic feature to build a \textbf{language-specific embedding} as Equation \ref{equ:concat}. For example, $[1, 0, 0,1,0]$ means that the user chooses both first and fourth languages at inference.


To further enhance the model ability of distinguishing different languages, we design a \textbf{language-specific layer} used in the encoder network or prediction network.
At layer $l$ of encoder network, we have the universal module (uni) and $N$ language-specific modules ($\texttt{Linear}_i$, $i=1...N$), where $N$ is the total number of languages in training, as shown in Figure \ref{fig:CMAT}. The layer input $v$ is passed into every module to generate the output $h_{uni}$ and $h_{spe,i}$. 
\begin{align}
    h_{att}^{l} &= \texttt{LayerNorm}(\texttt{Attention}(v^{l-1}) + v^{l-1}) \\
    h_{uni}^{l} &= \texttt{LayerNorm}(\texttt{FFN}(h_{att}^{l}) + h_{att}^{l}) \\
    h_{spe,i}^{l} &= \texttt{Linear}_{i}(h_{att}^{l}), i={1,2,...,N} \label{specific-layer-fun}
\end{align}
where \texttt{LayerNorm}, \texttt{Attention}, and \texttt{FFN} denotes layer normalization, self-attention, and feed-forward network, respectively.

Note that because we already have a universal module which models the shared information across all languages, it just needs much fewer parameters for each language-specific module to model the residue from the specific language. 
By combining the universal representation and specific representation, the formulation of the output at the $l$-th layer is
\begin{equation}
    v^l = h_{uni}^l + \sum_{i=1}^N w_i h_{spe,i}^l
\end{equation}


The weight $w_i$ is determined by the user choice vector:
\begin{itemize}
    \item 1-hot vector, i.e., the user selects only one language: $w_i$ will also be a 1-hot vector. 
    \item multi-hot vector, i.e., the user selects multiple languages: a vector with several 1 (corresponding to user choice) and all others 0. $w_i$ will be normalized by the total number of 1 in the vector.  
\end{itemize}

We further apply the specific module into the output of the prediction network.
Formally, by utilizing a feed-forward network, the joint network combines the encoder network output $h_t^{enc}$ and the prediction network output $h_u^{dec}$ as:
\begin{align}
    z_{t,u} &= f^{joint} (h_t^{enc}, h_u^{dec} ) \\
            &= \phi (U h_t^{enc} + V h_u^{dec} + \sum_{i=1}^N w_i h_{spe,i}^{dec} + b_z)
\end{align}
where $h_{spe,i}^{dec} = \texttt{Linear}_{i}(h_{u}^{dec})$, 
is the proposed language-specific prediction-network output for language $L_i$. $U$ and $V$ are weight matrices, $b_z$ is a bias vector, and $\phi$ is a
non-linear function, e.g., Tanh.


When deploying the user-specific model, we just need to extract out the corresponding language-specific module together with the universal module per user choice vector. 
Moreover, we design a \textbf{language-specific vocabulary} strategy. Given the vocabulary of each language ${V_1,...,V_N}$ and total vocabulary $V_{total}$, we can merge the corresponding vocabularies of user choice to a temporary vocabulary $V_{tmp}$ at inference. $V_{tmp}$ is smaller than $V_{total}$, which can be used to avoid the generation of unexpected tokens from other languages not selected by users.

\section{Training}
\label{sec:training}
In the multilingual scenario, given $N$ languages ${L_1,...,L_N}$, with its training set $\{(\mathcal{X}_1,\mathcal{Y}_1),...,(\mathcal{X}_N, \mathcal{Y}_N)\}$, the training loss for the model is minimizing the sum of the negative log probabilities over all training examples:
\begin{equation}
    \mathcal{L}(\theta) = - \sum^{N}_{n=1} \sum^{M_{n}}_{m=1} \sum^{U_m}_{u=0} log P(y_u^{m,n}|x_t^{m,n}, y_{u-1}^{m,n})
\end{equation}
where $M_{n}$ is the number of training examples in language $L_n$, and $U_m$ is transcription sequence length.

We have two strategies to train CMM. The first is to train CMM from scratch. The second one is to first train the universal module using the training data without user choice vector. Then we train language-specific modules using training data with user choice vector by fine-tuning the pre-trained model. To reduce memory consumption, we only apply a language-specific linear layer to the top and bottom layers instead of all encoder network layers, which doesn't require as many parameters as the universal module, and makes it very easy to scale up with multiple languages.

The key to train CMM is that we need to simulate the combination of languages selected by users. To do that, for each training sample, we generate the user choice multi-hot vector by randomly setting several (or zero for 1-hot vector) elements together with the ground truth element as 1, and setting other elements as 0. In this way, CMM is informed that the current training sample comes from one of the several languages set by the user choice vector. During training, we go through all the combinations of languages. 

\section{Experimental Setting}


\begin{table}[t]
\caption{Number of utterances in train and test sets.}
\label{tab:dataset}
\centering
\begin{tabular}{l|ll||l|ll}
\hline
LANG & Train & Test   & LANG & Train & Test  \\ \hline
EN    &  32.6M & 266.9K & ES    &  6.7M & 42.6K  \\
FR    &  5.9M  & 42.8K  & PT    &  3.6M & 21.4K \\
IT       &   6.0M                  & 24.7K  & NL       &  0.6M                  & 7.9K  \\
PL       &   1.4M                  & 6.1K   & DE       &  4.7M                  & 49.0K \\
RO       &   1.2M                  & 16.7K  & EL       &  1.5M                  & 26.0K \\ \hline
\end{tabular}
\end{table}

\begin{table*}[t]
  \caption{WER of baselines and our proposed configurable multilingual model.}
  \label{tab:CMM-M3}
  \centering
  \begin{tabular}{l|c|c|c||c|c|c}
    	\hline
			Language & Monolingual & Multilingual & Multilingual & CMM-M3 & CMM-M3 & CMM-M3        \\							
			         & Baseline & w/o 1-hot LID & w/ 1-hot LID & w/ 1-hot LID & w/ 2-hot LID  & w/ 3-hot LID \\
    	\hline
        EN  & 9.52    & 10.72  & 10.50  & 9.90   & 10.04   & 10.14  \\
        ES  & 19.98   & 19.83  & 16.07  & 14.82  & 15.88   & 17.04  \\
        FR  & 21.58   & 27.02  & 17.43  & 16.68  & 19.66   & 22.35   \\
        IT  & 19.67   & 21.59  & 15.30  & 12.57  & 14.65   & 16.60    \\
        PL  & 17.39   & 23.99  & 13.69  & 13.73  & 18.63   & 21.63   \\
        PT  & 14.58   & 14.14  & 13.01  & 12.26  & 12.86   & 13.40  \\
        NL  & 20.74   & 24.41  & 17.70  & 17.23  & 20.96   & 22.80    \\
        DE  & 16.26   & 18.16  & 16.24  & 15.44  & 16.46   & 17.18   \\
        RO  & 14.91   & 15.56  & 14.62  & 13.72  & 14.45   & 14.85   \\
        EL  & 17.63   & 17.83  & 17.43  & 16.57  & 16.98   & 17.20    \\
        \hline
        AVE & 17.22 & 19.32 & 15.20 & 14.29  & 16.06 & 17.32 \\
    	\hline
  \end{tabular}
\end{table*}

\subsection{Dataset}

We investigate the performance of the proposed configurable multilingual model on 75 thousand (K) hours of transcribed Microsoft data. The training set and test set cover 10 languages, including English (EN), Spanish (ES), French (FR), Italian (IT), Polish (PL), Portuguese (PT), Netherlands (NL), German (DE), Romanian (RO), and Greek (EL). The size of training data for each language varies due to the availability of transcribed data from 0.6 Million (M) utterances to 32.6M, which are shown in Table \ref{tab:dataset}. 
All the training and test data are anonymized data with personally identifiable information removed. 
Separate validation sets of around 5K utterances per language are used for hyperparameter tuning.
Besides, we use German/English (DE/EN) and Spanish/English (ES/EN) code-switching set to evaluate the ability of our model to address the code-switching challenge. In these two test sets, the majority of words in every utterance are German and Spanish, respectively, mixed with a few English words.

\subsection{Setting}

All experiments in this paper employ 80-dimensional log-Mel filter bank features, computed with a 25 millisecond (ms) window, and the frame shift is 10ms.
The features are normalized using global mean-variance statistics.
Following \cite{chen2020developing}, we apply a future context window of 18 and a left chunk of 4 for the input acoustic feature.
We set the vocabulary as 10K sentence pieces trained on the training data transcription of all languages.
Data sampling was applied to solve the data imbalance issue of multilingual corpus.
In terms of Transformer Transducer, 18 transformer layers with 512 hidden units and 2048 feed-forward nodes are used as the encoder network, and 2 LSTM layers  with 1024 memory cells are used as the prediction network.
Finally, the joint network also has 512 hidden units.
We also use the relative position encoding to boost the modeling of position information.

The language specific module ($\texttt{Linear}_i$) is a linear layer, consisting of a parameter metrix $W$ $\in$ $\mathbb{R}^{512 \times 512}$ for each language in the encoder network.
A 10-dimensional multi-hot language vector is fed into the encoder as an additional input to CMM. 
The Adam algorithm \cite{loshchilov2017decoupled} with gradient clipping and warmup is used for optimization.
All the transducer models used in this paper are implemented with Pytorch.
We train models using 32 NVIDIA V100 GPUs, and report the word error rate (WER) for every language and also averaged WER over all languages.

\subsection{Models}


We list the three baselines and two CMMs below, all of which are based on Transformer Transducer model architectures.

\begin{itemize}
	\item \textbf{Monolingual baseline}: As a baseline, we train ten monolingual models independently on the data from each language.
	\item \textbf{Multilingual w/o LID baseline}: It is a universal multilingual model without LID, which is trained with the same model architecture as monolingual models, but with training data combined from all languages.
	\item \textbf{Multilingual w/ 1-hot LID baseline}: It is a multilingual model with LID, which concatenates a given 1-hot LID vector to the input features with training data from all languages, as introduced in Section \ref{multilingual-speech-recognition}.
	\item \textbf{CMM-M3}: It is a CMM that supports monolingual, bilingual, and trilingual combinations of 10 languages, namely CMM-M3 w/ 1-hot LID, CMM-M3 w/ 2-hot LID, and CMM-M3 w/ 3-hot LID.
	\item \textbf{CMM-M10}: To evaluate the scalability of our model, we train another CMM, which allows users to select up to 10 languages.
\end{itemize}

The numbers of parameters of multilingual w/o LID baseline, multilingual w/ 1-hot LID baseline, and CMM are 80.9M, 81.0M, and 91.5M, respectively. Our proposed CMM only increases 13\% parameters compared to the multilingual model w/o LID, and the increased parameters are mainly from the linear layers of both the encoder network and the prediction network.

\section{Results}


\subsection{Multilingual vs. monolingual models}

We first compare the monolingual models and multilingual models. As shown in Table \ref{tab:CMM-M3}, compared to the monolingual baselines, the universal multilingual model without LID which simply concatenates training samples of all languages 
gets a relative 12.2\% WER increase on average for all 10 languages. 
These results show the challenge of the universal model without knowing which language the user will speak in advance. 

If knowing which language the user will speak in advance, the multilingual model with 1-hot LID can achieve a 21.3\% relative WER (WERR) reduction from the multilingual universal model without LID, and it outperforms monolingual baselines by 11.7\% WERR over ten languages.
The experimental results show the importance of leveraging user selection by taking a 1-hot LID vector as the additional input that can guide the recognition of the current language.

\begin{figure}
  \centering
  \includegraphics[width=0.95\linewidth]{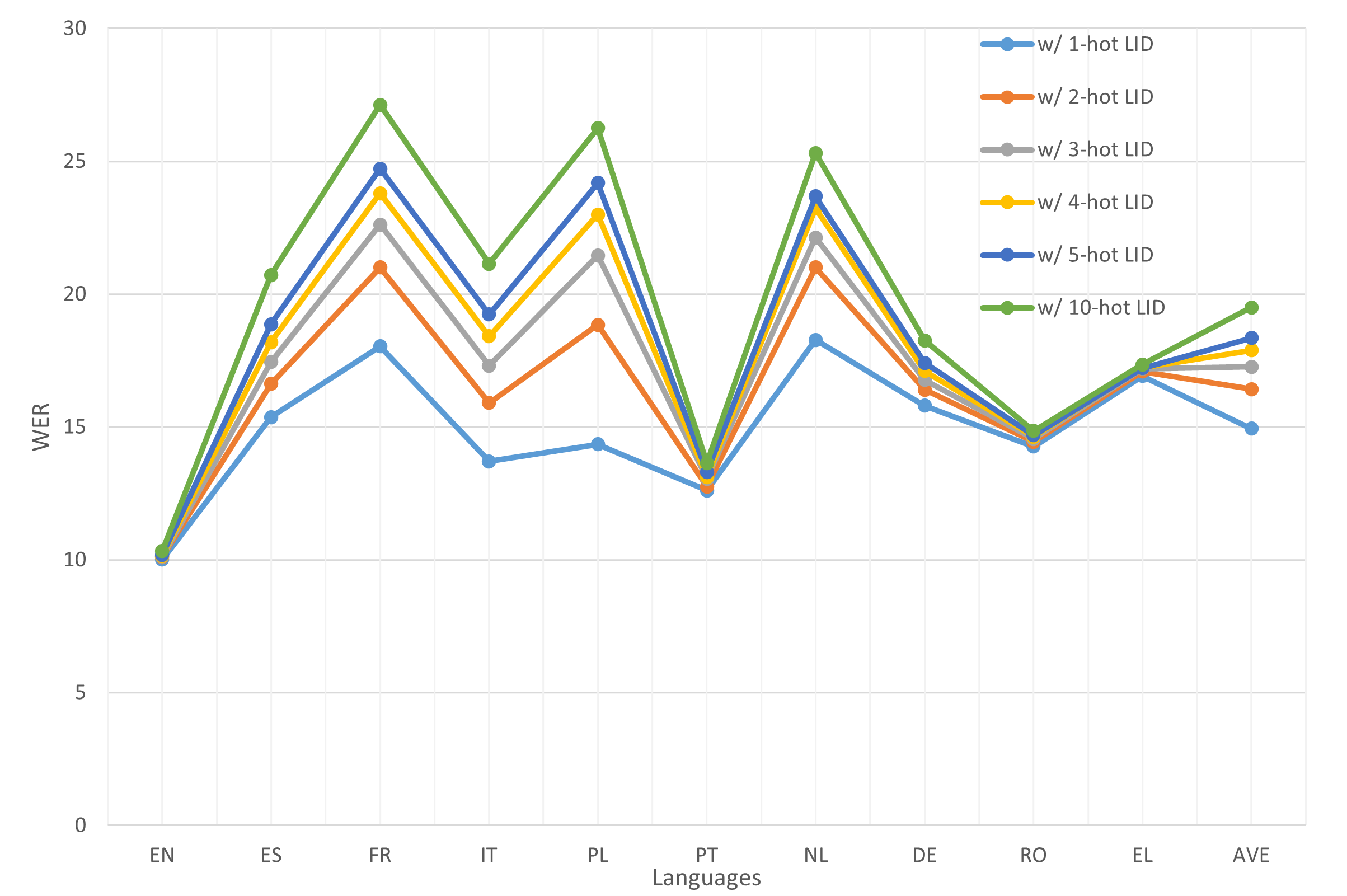}
  \caption{Average WER of configurable multilingual model with different multi-hot vector on 10 languages (CMM-M10).}
  \label{fig:CMM-M10}
\end{figure}

\subsection{CMM vs. monolingual/multilingual models}

The results of CMM-M3 are also shown in Table \ref{tab:CMM-M3}.
Similar to the multilingual model with 1-hot LID, CMM-M3 with 1-hot LID has the same setting at inference and achieves better performance (6.0\% WERR) than the multilingual model with 1-hot LID, which demonstrates that the proposed language-specific modules are beneficial for multilingual speech recognition.
CMM-M3 also supports bilingual and trilingual languages decoding. During evaluation, in addition to the current language, we also randomly assign one or two other languages by constructing 2-hot and 3-hot LID vectors to simulate the cases that users select two or three languages, respectively. 
The results show that the performance of our proposed CMM-M3 with 2-hot LID and CMM-M3 with 3-hot LID is between the universal multilingual model without LID and the specific multilingual model with 1-hot LID, which meets our experimental expectations of supporting user's multiple selections: supporting more languages brings more confusion which reduces the model's focus on a single language. CMM improves from the universal multilingual model by 26.0\%, 16.9\%, and 10.4\% WERR when the user selects 1, 2, or 3 languages, respectively.
For a fair comparison, we also enlarge the hidden layer size of the universal model to have a similar parameter size as the CMM. Results show that the enlarged universal model gets a slight gain than the standard universal model (18.83 vs. 19.32), and our CMM still significantly outperforms the enlarged universal model.

We further conduct the experiments of CMM-M10, in which any combination of 10 languages can be chosen by users. 
At inference, we verify the configurable model given 1-hot, 2-hot, 3-hot, 4-hot, 5-hot, or 10-hot LIDs.
As shown in Figure \ref{fig:CMM-M10}, we can draw several conclusions: (1) The CMM training method is still effective when the maximum combination is expanded to all 10 language; (2) The more languages the user selects at inference, the higher the WER of the CMM; (3) CMM-M10 with 10-hot LID can recognize the same 10 languages as the universal model, both of which achieve comparable performance on all languages; (4) Due to more user choices, the frequency of each language combination in training will decrease. This is a potential reason that CMM-M10 performs worse than CMM-M3 on the cases with 1-hot, 2-hot, and 3-hot LID, while CMM-M3 cannot handle the recognition with more than 3 languages selected by users.

\begin{table}[t]
\caption{WER of universal model and configurable model on code-switching corpus. }
\centering
\begin{tabular}{l|llll}
\hline
CorpusName & UttCount  & Baseline & CMM  & WERR    \\
\hline
DE/EN      & 1996      & 36.39          & 34.63  &  4.8\%          \\
ES/EN      & 1827     & 27.29          & 22.85   &  16.3\%        \\ 
\hline
\end{tabular}
\label{tab:code-switching}
\end{table}

\begin{table*}[t]
  \caption{Ablation study. The CMM-M3 with 2-hot LID, which is trained from scratch and decoded with 2-hot LID, is used as baseline. ``- specific embedding", ``- specific layer", and ``- specific vocabulary" means removing the language-specific embedding, layer, and vocabulary, respectively. ``- encoder SL" and ``- prediction SL" means the CMM without specific layer in encoder network or prediction network. CMM-M3-Finetune with 2-hot LID denotes the model fine-tuned from the universal model  and decoded with 2-hot LID.}
    \label{tab:ablation-study}
  \centering
  \begin{tabular}{l|c||c|c|c||c|c||c}
    \hline
    Language & \begin{tabular}[c]{@{}c@{}}CMM-M3 \\ with 2-hot LID\end{tabular} & \begin{tabular}[c]{@{}c@{}}- Specific\\ embedding\end{tabular} & \begin{tabular}[c]{@{}c@{}}- Specific\\ layer\end{tabular} & \begin{tabular}[c]{@{}c@{}}- Specific\\ vocabulary\end{tabular} & \begin{tabular}[c]{@{}c@{}}- Encoder\\ SL\end{tabular} & \begin{tabular}[c]{@{}c@{}}- Prediction \\ SL\end{tabular} &  \begin{tabular}[c]{@{}c@{}}CMM-M3-Finetune \\ with 2-hot LID\end{tabular}  \\
    \hline
    EN  & 10.04   & 10.10  & 10.50   & 10.04  & 9.99   & 10.52  & 9.43    \\
    ES  & 15.88   & 16.26  & 16.91   & 15.88  & 16.57  & 16.66  & 15.80    \\
    FR  & 19.66   & 20.62  & 21.79   & 19.67  & 20.87  & 20.68  & 21.36  \\
    IT  & 14.65   & 14.78  & 15.24   & 14.67  & 14.56  & 14.79  & 14.50    \\
    PL  & 18.63   & 18.68  & 19.09   & 18.65  & 18.94  & 18.79  & 19.29   \\
    PT  & 12.86   & 12.80  & 13.60   & 12.86  & 13.18  & 13.26  & 11.88   \\
    NL  & 20.96   & 20.83  & 21.77   & 20.95  & 21.90  & 22.10  & 20.11   \\
    DE  & 16.46   & 16.45  & 16.84   & 16.46  & 16.27  & 16.92  & 15.63   \\
    RO  & 14.45   & 14.36  & 15.19   & 14.45  & 14.48  & 15.08  & 13.42   \\
    EL  & 16.98   & 17.09  & 17.51   & 16.98  & 17.08  & 17.59  & 16.11   \\
    \hline
    AVE & 16.06 & 16.20 & 16.84 & 16.06 & 16.38 & 16.64 & 15.75 \\
    \hline
  \end{tabular}
\end{table*}

\subsection{Results on code-switch corpus}

In this section, we evaluate the proposed configurable multilingual model on the code-switching task. Since CMM can support bilingual speech recognition, it is a potential function to tackle the code-switching problems. Table \ref{tab:code-switching} lists the experimental results on German/English and Spanish/English code-switching datasets.
The baseline model is the universal multilingual model, which obtains 36.39\% and 27.29\% WER DE/EN and ES/EN dataset, respectively. We use CMM-M3 with 2-hot LID as our bilingual configurable model, which outperforms the universal model by 4.84\% and 16.30\% WERR, respectively. The significant improvement demonstrates the validity of our proposed CMM on code-switching tasks. Note that, unlike the previous work \cite{li2019towards, zhou2020multi,winata2020meta,dalmia2021transformer}, we do not make 
other specific model design for this code-switching task.

\subsection{Ablation study}

Different from the conventional multilingual model, the proposed CMM employs three specific modules, including language-specific embedding, language-specific layer, and language-specific vocabulary, as introduced in Section \ref{sec:cmm}. In this section, we first conduct an ablation study to analyze the effectiveness of each module.

We show in Table \ref{tab:ablation-study} the WER performance of different configurable model variants for CMM-M3 with 2-hot LID task. The CMM without language-specific embedding and layer obtains the average performance of 16.20\% and 16.84\% WER on all languages,  worse than baseline CMM by 0.8\% and 4.8\% relative, respectively. It demonstrates that language-specific linear layer is more important than language-specific embedding for CMM. Although the CMM without language-specific vocabulary obtains similar WER, employing language-specific vocabulary can avoid outputting the unexpected token of languages not selected by users, hence improving user experience. 


Second, as in the previous analysis, the language-specific layer is the key component of the proposed CMM. This prompts us to further break down the contribution of the language-specific layer into the contribution from the encoder network and the prediction network.
The results of removing the language-specific layer in the encoder and prediction network are listed in the six and seven columns of Table \ref{tab:ablation-study}, in which CMM without the language-specific layer of the prediction network perform worse than CMM without the language-specific layer of the encoder.
Therefore, the specific layer in the prediction network is slightly more critical than the specific layer in the encoder network in CMM.

Finally, we compare two different training methods: training from scratch and fine-tuning from a universal model, as introduced in Section \ref{sec:training}.
CMM-M3 is trained from scratch, and CMM-M3-Finetune uses the same setting but it is fine-tuned from a universal model. The two models get 16.06\% and 15.75\% average WER, respectively, which demonstrates that the fine-tuning strategy is better than training from scratch for CMM.

\section{Conclusion}

In this paper, 
we proposed a configurable multilingual model (CMM) which consists of a universal multilingual module and a specific module for each language.
Through our designed training algorithm, CMM can be configured to recognize speeches from any combination of languages while taking advantage of user selection, which means that the single CMM can be applied to any usage scenario.
More importantly, we only train CMM once but can deploy different models based on user choice by using language-specific embedding, layer, and vocabulary. Because most language information is modeled by the universal multilingual model, the language-specific layer is small to model the language residue and hence CMM is only slightly larger than the universal multilingual model. 

Massive experiments are conducted on 75K hours of transcribed anonymized Microsoft data with 10 languages. After user's language selection, both CMM and the universal multilingual model without LID don't need to know in advance which language the user will speak, while the multilingual model with 1-hot LID has to know. Results demonstrate that CMM improves from the universal multilingual model without language information by 26.0\%, 16.9\%, and 10.4\% WERR when the user selects 1, 2, or 3 languages, respectively. The improvement on two code-switching tasks is 4.8\% and 16.3\% WERR, respectively.

\bibliographystyle{IEEEbib}
\bibliography{main}

\end{document}